\magnification1200

\rightline{KCL-MTH-04-09}
\rightline{hep-th/0407088}

\vskip .5cm
\centerline
{\bf  Some simple predictions from  $E_{11}$ symmetry}
\vskip 1cm
\centerline{Peter West}
\centerline{Department of Mathematics}
\centerline{King's College, London WC2R 2LS, UK}

\vskip 0.5cm
\centerline{and}
\vskip 0.5cm
\centerline{Erwin Schr{\" o}dinger International Institute for
Mathematical Physics,}
\centerline {Boltzmanngasse 9,} 
\centerline{1090 Wien, Austria}
\leftline{\sl Abstract}
\vskip .2cm
\noindent 
The   simplest  consequences  of the   common $E_{11}$ symmetry of the
eleven dimensional, IIA and IIB theories are derived and   are shown
to imply  the known relations between these  three  theories. 

\vskip .5cm

\vfill
\eject
\medskip
{\bf {1. Introduction }}
\medskip
It has been argued that eleven dimensional supergravity [1]
when suitably extended possess a 
non-linearly realised $E_{11}$ symmetry  [2]. Furthermore,
 both  the IIA supergravity [3] and IIB
supergravity  theories [4,5], when suitably extended, 
 should also possess a  non-linearly realised
$E_{11}$ symmetry [2,6].  As explained below,  the three different
theories arise from the same underlying algebra due to the different
possible embeddings in $E_{11}$ of the sub-algebras that  describe 
gravity [2,6]. Their 
common
$E_{11}$ symmetry can be exploited to find explicit relations between the
eleven dimensional, IIA and IIB non-linearly realised theories [7].
Indeed, one can find a one to one correspondence between  the fields that
occur in any two of these three theories providing a very concrete idea
of what M theory actually is [7].  In this paper we derive the simplest
of these  relations which are those that involve fields that are
associated with the Cartan sub-algebra of
$E_{11}$. We recover the known relations between the 
eleven dimensional, IIA and IIB  theories, when dimensionally reduced on a
suitable torus,  in a simple way. Originally these relations were found by
using a mixture of  string and  solitonic properties [8,9,10,11,12,13],
but we will show that they follow from the way the sub-algebra
associated with the gravity sectors of the different theories are embedded
in $E_{11}$. We also give an example of how the correspondence works for
a field not associated with the Cartan sub-algebra and derive the
effect of the Weyl transformations of the $E_8^{+++}$ theory for the
   IIA and IIB theories. 
\par
A  Kac-Moody
algebra  is specified by its  Cartan matrix 
$A_{ab}$  which by definition must   satisfies the following properties:  
$$A_{aa}=2, 
\eqno(1.1)$$ 
$$A_{ab}\  {\rm for}\  a\not= b\  {\rm are\  negative\  integers\  or
\ zero}, 
\eqno(1.2)$$
and  
$$  A_{ab}=0\  {\rm implies}\  A_{ba}=0 
\eqno(1.3)$$ 
The index range $a,b=1,\ldots, r$ where $r$ is the rank of the
Kac-Moody algebra. The   Kac-Moody  algebra is  formulated in terms of its
Chevalley generators denoted  by 
$H_a, E_a, F_a$, $a=1,\ldots, r$ which  obey the Serre  relations;  
$$[H_a, H_b]= 0, 
\ \ [H_a, E_b]= A_{ab} E_b, \ \ 
[H_a, F_b]= -A_{ab} F_b, 
\ \ [E_a, F_b]= \delta_{ab} H_a, 
\eqno(1.4)$$
and  
$$[E_a,\ldots [E_a, E_b]\ldots ]= 0, \  
[F_a,\ldots [F_a, F_b]\ldots]= 0
\eqno(1.5)$$ 
In equation (1.5)  there are
$1-A_{ab}$ 
$E_a$'s in  the first equation and the same number of $F_a$'s in the
second equation.  Given the Cartan matrix 
$A_{ab}$,  one can construct the 
Kac-Moody  algebra by taking
the multiple  commutators of the $E_a$'s, and separately the $F_a$'s, 
subject to the above Serre relations. We recognise the 
$H_a$ generators as those of the   Cartan  sub-algebra and 
the $E_a$'s
($F_a$'s) as the generators  of the positive (negative) simple
roots,   hence, a Kac-Moody algebra is uniquely specified by its 
 Cartan matrix $A_{ab}$ which is also encoded in its Dynkin
diagram. For the case when all the simple roots are of equal length, the
Dynkin diagram  consists of $r$ dots  labeled $a=1,\ldots , r$
with
$-A_{ab}$ lines between nodes labeled $a$ and $b$. 
The Cartan matrix can be expressed in terms of the
simple roots 
$\alpha_a$ as 
$A_{ab}=2{(\alpha_a, \alpha_b)\over (\alpha_a,\alpha_a)}$. 
Although the construction of the Kac-Moody algebra is simple in theory,
in practice it is difficult and the generators are not known explicitly
for any Kac-Moody algebra except those that are affine, finite
dimensional or for certain special algebras associated with string
theory. 
\par
We invite the reader to
draw the Dynkin diagram of $E_{11}$ by drawing ten nodes  connected
together by a single horizontal  line. We label these
nodes  from left to right  by the integers  one to ten and then add a
further node, labeled eleven, above node eight and attached by
a single  vertical line. This algebra is  just one of a special class of
algebras, called very extended algebras, which were studied in [14]. One
begins with the Dynkin diagram of any finite dimensional semi-simple lie
algebra
$G$, and  adds the affine node to find the affine algebra $G^{+}$. One
then adds a  further node (called the extended node)  attached to the
affine node by a single horizontal line and finally yet another node
(called the very extended node) attached to the extended  node by a
single horizontal line. One denotes the very extended algebra by
$G^{+++}$. Hence, we also denote  $E_{11}$ as  $E_8^{+++}$. 
\par
As the reader may readily verify, any Kac-Moody  algebra is invariant
under the Cartan involution which acts on the Chevalley generators as 
$E_a \to - F_a,\  F_a \to - E_a,\ 
H_a\to -H_a$. 
The sub-algebra that is invariant under the Cartan involution is then
generated by
$E_a  - F_a$.   We note that this sub-algebra contains none of
the   generators $H_a$ 
of  the Cartan sub-algebra of the Kac-Moody algebra. 
\par
We now give a brief exposition of non-linear realisations so that the
reader may appreciate the more general setting, although much of this is
not required later on in the paper. 
A non-linear realisation is specified by an algebra together with a
chosen local sub-algebra. In the case of interest in this paper these
are $E_8^{+++}$ and, essentially, the Cartan invariant sub-algebra
respectively. The non-linear realisation is then just a theory which is
invariant under $g\to g_0 g h$ where $g$ and $g_0$ belong to the group
and $h$ to the local  subgroup and also $g$ and $h$ are local. The
meaning of local  depends on how one encodes space-time and we refer the
reader to reference [19] for a recent discussion of this point for the
theories being considered here. However, in the conventional applications
of non-linear realisations, and in essence also here,  it means that $g$
and
$h$, unlike $g_0$, depend on space-time. In general these invariances are
not sufficient to determine the theory uniquely, but if the local
sub-algebra is large enough it they do.  The local transformations  allow
one to choose  
$g$ in the form
$g=\exp (\sum A\cdot R)$ where the sum is over all generators of the
algebra that are not in the local sub-algebra and their coefficients are
the fields of the theory.  In our case, $g$ then contains all the
generators of the Borel sub-algebra of $E_8^{+++}$. 
\par 
So little is
known about Kac-Moody algebras that it is difficult to calculate the
general properties of a non-linear realisation based upon  them. 
However, by setting to zero all the fields of the non-linear realisation,
except those associated with the  Cartan 
sub-algebra, the group element takes the very simple form 
$$
g=\exp(\sum_a q_a H_a)
\eqno(1.6)$$
The fields $q_a$ are then the only  fields of the theory. 
 Provided one restricts ones attention to operations  that preserve
the Cartan sub-algebra  it is then essentially trivial to examine the
consequences.  Such is the case for Weyl transformations. Indeed, these
were considered  in just such a setting for the 
$E_8^{+++}$ non-linear realisation  appropriate to the eleven dimensional
theory and the Weyl  transformations  were shown [15] to be are none other
than  the U-duality transformations [18]. 
\par
The algebra $G^{+++}$ contains a GL(D) sub-algebra, with
generators
$K^a{}_b$, $a,b=1,\ldots, D$ 
which leads  in the non-linear  realisation to the  gravity sector of
the resulting theory where $D$ is the dimension of the space-time of
the theory. The $A_{D-1}$, or SL(D), part of this  sub-algebra is obtained
by taking
$D-1$ dots of the Dynkin diagram of $G^{+++}$ which are connected to
the very extended node, i.e. selecting an $A_{D-1}$ sub-Dynkin diagram
which contains as an extreme node the very extended node.  As we shall
see,  there is more than one way to do this in general and these lead to
different physical theories. The part of the group element of
$G^{+++}$ which contains the generators of the preferred sub-algebra  is
of the form 
$\exp(\sum_{a\le b}h_a{}^b K^a{}_b)$ and carrying out the non-linear
realisation one finds that   the vierbein 
$e_\mu{}^a$ is identified with $e_\mu{}^a=(e^h)_\mu{}^a$,  where in
this last equation, we treat $h$ as a matrix [2,16]. This Gl(D), or in
some cases the SL(D), 
 sub-algebra is referred to as the gravity sub-algebra and the 
$D-1$ dots of the Dynkin diagram of $G^{+++}$ which belong to the  SL(D),
or $ A_{D-1}$, sub-algebra are
referred to  as the gravity line. 
\par
The eleven dimensional, IIA and IIB theories all have an underlying 
$E_8^{+++}$, but they are distinguished by their different gravity
sub-algebras. The eleven dimensional theory must possess an 
$A_{10}$ gravity algebra  and there is only one such algebra.
We must choose the  gravity algebra to be  the $A_{10}$ sub-Dynkin diagram
that consist of nodes labeled one
to ten. That is it is found by deleting node eleven in the
$E_8^{+++}$  Dynkin diagram [2].
\par
The IIA and IIB theories are ten dimensional and so to find 
these theories we seek an $A_9$ gravity algebra. Looking at the
$E_8^{+++}$ Dynkin diagram there are only two ways to do this.  Starting
from the very extended node  we must choose a $A_9$ sub-Dynkin diagram,
but once we get to  the
junction of $E_8^{+++}$ Dynkin diagram, situated at the node labeled 8,
 we can continue along the horizontal line 
with two further nodes taking only the
first node to belong to the $A_9$, or we can find the final $A_9$ node by
taking it to be the only node  in the other choice of direction at the
junction.  These two ways correspond to
the IIA and IIB theories respectively. Hence, in the IIA theory we take
the gravity line to be nodes labeled one to nine inclusive while for the
IIB theory the gravity line contains nodes one to eight and in addition
node eleven [2,6]. 
\par
The gravity  sub-algebra is such that $K^a{}_a, \ a=1,\ldots, D$
are part of the Cartan sub-algebra of $E_8^{+++}$. 
For the 
eleven dimensional theory,  these eleven generators  span the Cartan
sub-algebra  and so one can also write the group element of equation (1.6)
in the form 
$$g=\exp (\sum_{a=1}^{11} h_a{}^a K^a{}_a)=\exp (h^T K)
\eqno(1.7)$$
In the second equation  we have used matrix
notation whose meaning should be clear. The  relationship  between the
Chevalley generators 
$H_a$ and the physical generators 
$K^a{}_a$ can be written  in matrix form as $K=\rho H$. It is 
given by  [2] 
 $$ H_a= K^a{}_a-K^{a+1}{}_{a+1}, a=1,\ldots ,10, \ 
H_{11}=
-{1\over 3}(K^1{}_1+\ldots +K^8{}_8)+{2\over 3}(K^9{}_9
+K^{10}{}_{10}+K^{11}{}_{11}). 
\eqno(1.8)$$
We also record the relations 
$$E_a=K^a{}_{a+1}, a =1, \ldots 10, \ 
E_{11}= R^{91011},
\eqno(1.9)$$
 between the Chevalley
generators $E_a$ and the simple root generators of SL(11) and the
generator
$R^{a_1a_2a_3}$ which is responsible in the non-linear realisation for the
introduction of the gauge field $A_{a_1a_2a_3}$ of the eleven dimensional
supergravity theory. Hence, keeping only fields associated with the Cartan
sub-algebra implies keeping only the diagonal parts of the metric and, as
we will see below for the IIA and IIB theories, also the dilaton field. 
\par
The form of the $H_a$ of equation (1.8) can essentially be   determined 
given that they must obey equation (1.4) with the Cartan matrix of  
$E_8^{+++}$, together with the knowledge of the simple roots generators 
of equation (1.10)  and that tensors, such as $R^{a_1a_2a_3}$, transform
in the obvious way under   GL(11), i.e. 
$[K^c{}_d, R^{a_1a_2a_3}]=\delta^{a_1}_d R^{c a_2a_3}
+\delta^{a_2}_d R^{c
a_3a_1}+\delta^{a_3}_d R^{c a_1a_2}$.  
\par
We denote quantities in the IIA and IIB theories with a tilde and hat
respectively. For these theories the Cartan sub-algebra of the
gravity sub-algebra, i.e the  $\tilde K^a{}_a, \ a=1,\ldots, 10$ for the
IIA theory and the 
$\hat K^a{}_a, \ a=1,\ldots, 10$ for the
IIB theory,  account for only ten of the eleven generators of
the Cartan sub-algebra of
$E_8^{+++}$. The final commuting generator is associated with the 
dilaton which appears in the IIA and IIB
theories. We denote this generator 
 by the symbol $R$ and the dilaton by $A$ with appropriate tildes or
hats. As such, for the IIA theory the 
$E_8^{+++}$ group element of equation (1.6) can be written in terms of
the physical generators in the form 
$$\tilde g=\exp (\sum_{a=1}^{10} \tilde h_a{}^a \tilde K^a{}_a)\exp(\tilde
A\tilde R) =\exp (\tilde h^T \tilde K)
\eqno(1.10)$$
In the second equation we have used matrix notation for which  
$\tilde h$ is a column vector whose first ten components are
$\tilde h_a{}^a,\ a=1,\ldots ,10$ and whose eleventh component is $\tilde
A$ and similarly
$\tilde K$ has its first ten components as $\tilde K^a{}_a$ and eleventh
component $\tilde R$. 
The Cartan
sub-algebra generators $H_a$ of
$E_{11}$ and the physical generators $\tilde K^a{}_a, \ a=1,\ldots, 10$ 
and $\tilde R$  are related by  $H=\tilde  \rho \tilde  K$ which is given
by [2]
$$ H_a= \tilde K^a{}_a-\tilde K^{a+1}{}_{a+1}, a=1,\ldots ,9,
\  H_{10}= -{1\over 8}(\tilde K^1{}_1+\ldots +\tilde K^9{}_9)+{7\over
8}\tilde K^{10}{}_{10} -{3\over 2}\tilde R,
$$
$$  H_{11}= -{1\over 4}(\tilde K^1{}_1+\ldots +\tilde K^8{}_8)+{3\over
4}(\tilde K^9{}_9 +\tilde K^{10}{}_{10})+\tilde R. 
\eqno(1.11)$$
While the
$E_a$ Chevalley generators of
$E_8^{+++}$  are given in terms of IIA generators  by [2] 
$$E_a=\tilde K^a{}_{a+1},\ a =1, \ldots, 9, \  E_{10}= \tilde R^{10},\ 
E_{11}=\tilde  R^{910}.   
\eqno(1.12)$$
The fields associated with the generators $\tilde R^{a}$ and 
$R^{ab}$ in the non-linear realisation are the one form and two form
fields of the IIA supergravity theory. 
\par
Equating the Chevalley generators $H_a$ of equations (1.8) and (1.11) 
we find that the generators in the physical basis of the eleven
dimensional and IIA theory are related by [2] 
$$
K^a{}_a=\tilde K^a{}_a,\  a =1, \ldots, 10,\ 
K^{11}{}_{11}={1\over 8}\sum_{a=1}^{10}\tilde  K^a{}_{a}+{3\over 2}\tilde
R
\eqno(1.13)$$
\par 
For the IIB theory, the  generators  $\hat K^a{}_a, \ a=1,\ldots,
10$ 
and $\hat R$ span the Cartan sub-algebra of $E_8^{+++}$ and so the group
element of equation (1.6) can be expressed as 
$$\hat g=\exp (\sum_{a=1}^{10} \hat h_a{}^a \hat K^a{}_a)\exp(\hat  A\hat
R) =\exp (\hat h^T \hat K)
\eqno(1.14)$$
In the second equation we have used matrix notation for which  
$\hat h$ is a column vector whose first ten components are $\hat h_a{}^a,\
a=1,\ldots ,10$ and whose eleventh component is $\hat A$ and similarly
$\hat K$ has its first ten components as $\hat K^a{}_a$ and eleventh
component $\hat R$.
The relationship between the  Cartan sub-algebra generators $H_a$  of
$E_{8}^{+++}$ and the physical generators $\hat K^a{}_a, \ a=1,\ldots,
10$ and $\hat R$  can be written in the form  $H=\hat \rho \hat
K$ and it is explicitly given by [6]
$$
H_a=\hat K^a{}_a - \hat K^{a+1}{}_{a+1}, a=1,\ldots, 8,\
H_{9}=\hat K^{9}{}_{9}+\hat K^{10}{}_{10}+\hat R-{1\over
4}\sum_{a=1}^{10}\hat K^a{}_a, 
$$
$$
   H_{10}=-2\hat R,\ H_{11}=\hat K^{9}{}_{9}-\hat K^{10}{}_{10}
\eqno(1.15)
$$ 
The Chevalley generators $E_a$ of $E_8^{+++}$,  as they  appears in IIB
theory  are given by [6] 
$$
E_a=\hat K^a{}_{a+1}, a=1,\ldots 8,\  E_9=\hat R_1^{9 10},\
E_{10}=\hat R_2,\  E_{11}=\hat K^9{}_{10}.
\eqno(1.16)
$$
The fields associated with the generators $\hat R_1^{a b}$ and 
$\hat R_2$ are the NS-NS two form and the axion, $\hat \chi$ of the IIB
theory. 
The last equation 
reflects the fact that the node labeled eleven is the last node in the
IIB gravity line. 
\par 
Equating the Chevalley generators $H_a$ of equations (1.8) and (1.15) 
we find that the generators in the physical basis of the eleven
dimensional and IIB theory are related by [7] 
$$
K^a{}_a=\hat K^a{}_a,\  a =1, \ldots, 9, 
\hat K^{10}{}_{10}={1\over 3}\sum_{a=1}^{9} K^a{}_{a}
-{2\over 3}(K^{10}{}_{10}+K^{11}{}_{11}),\ 
\hat R=-{1\over 2}(K^{10}{}_{10}-K^{11}{}_{11})
\eqno(1.17)$$
\par 
For completeness we note the relationship between the IIA and IIB
physical generators;
$$
\hat K^a{}_a=\tilde K^a{}_a,\  a =1, \ldots, 9,\ 
\hat K^{10}{}_{10}={1\over 4}\sum_{a=1}^{9} \tilde K^a{}_{a}
-{3\over 4}\tilde K^{10}{}_{10}-\tilde R,\ 
$$
$$
\hat R={1\over 16}\sum_{a=1}^{9} \tilde K^a{}_{a}
-{7\over 16}\tilde K^{10}{}_{10}+{3\over 4}\tilde R,
\eqno(1.18)$$
\par
We note that the generator corresponding to  the node labeled ten
in the eleven dimensional theory is $K^{10}{}_{11}$ and so is
associated with the exchange of the ten and eleven space-time
coordinates, while in the IIB theory it is $\hat R_2$ which is the
non-perturbative part of the SL(2,{\bf Z}) symmetry of the IIB theory.

\medskip
{\bf {2. Relations between the eleven dimensional, IIA and  IIB theories}}
\medskip
As explained in reference [7], the common $E_8^{+++}$  origin of
these three theories implies a one to one correspondence between the
  fields  of the three theories.  In particular, any   field in
the non-linearly realised IIB theory arises in the group element as the
coefficient of  a
particular generator which is  in the Borel sub-algebra of $E_8^{+++}$,
however,  the  generators of $E_8^{+++}$ are essentially  unique and so we
can identify this generator 
from the viewpoint of the eleven dimensional theory. 
For example, the component  graviton field  $\hat h_{9}{}^{10}$ of the
 IIB theory is  associated with the generator $\hat K^{9}{}_{10}$
which is equal to the Chevalley generator $E_{11}$ of $E_8^{+++}$.
However, from the eleven dimensional perspective this Chevalley generator
is equal to the generator $R^{91011}$ that  is associated with the
field  $A_{91011}$ which is one component of the   third rank
anti-symmetric field of the eleven dimensional supergravity theory. 
In this section, we will find these correspondences at the simplest
possible level. 
\medskip
{\bf {2.1 The correspondence between the eleven dimensional and  IIA
theories}}
\medskip
To find the correspondence  for the Cartan sub-algebra we
simply equate the two group elements in the eleven dimensional and IIA
theories of equations (1.7) and (1.10) respectively;
$$g=\tilde g\ \ {\rm or }\ \ \exp (\sum_{a=1}^{11} h_a{}^a K^a{}_a)=\exp
(\sum_{a=1}^{10}
\tilde h_a{}^a
\tilde K^a{}_a)\exp(\tilde A\tilde R).
\eqno(2.1)$$
Using equation (1.13), we conclude that 
$$\tilde h_a{}^a= h_a{}^a+{1\over 8}h_{11}{}^{11}, \  a=1,\dots , 10, \ 
\tilde A={3\over 2} h_{11}{}^{11}
\eqno(2.2)$$
We expect these relations to hold even if one does not carries out 
dimensional reduction of the theory on a torus, but then one must also 
carry out a corresponding exchange of the generalised coordinates [17].
However, if we do dimensionally reduce some of the dimensions on a torus
then it is useful to change to the variables  
$$
h_a{}^a=\ln {R_a\over l_p},\ a=1,\dots , 11
\eqno(2.3)$$
where $l_p$ is the eleven dimensional Planck scale. We note that in the
group elements used to construct the non-linear realisation the fields
are dimensionless and so the resulting part of the action in $D$
space-time dimensions that has two space-time derivatives  is multiplied by
$l_p^{-(D-2)}.$  In particular, we will apply the change of variable to
the constant background part of the fields.  For a rectangular torus,
the coordinate and parameterisation invariant length of its cycle in the a
direction is
$l_p\int e_{a}{}^{a} d x^a=R_a$. 
\par
Similarly we introduce the
analogous IIA variables by 
$$
\tilde h_a{}^a=\ln {\tilde  R_a\over \tilde l_p},\ a=1,\dots , 10,\ \ 
\tilde A=\ln \tilde g_s
\eqno(2.4)$$
where $\tilde l_p$ is the ten dimensional Planck scale of the IIA theory.
Comparing the
low energy action with that calculated from string scattering allows
us to identify the string scale $l_s$ by $(\tilde l_p)^8=\tilde g_s^2
(\tilde l_s)^8$ and $\tilde g_s$ in equation (2.4) with the string
coupling constant in the usual way. 
\par
The last relation in equation (2.2) implies that 
$$(\tilde g_s)^2=\left({R_{11}\over l_p}\right)^3
\eqno(2.5)$$
Since the  eleven dimensional theory   after reduction on
a circle coincides with the IIA theory we may take $\tilde R_a=R_a,\
a=1,\dots , 10$  and then we find that 
$$\left({l_p\over \tilde l_p}\right)^{12}=\tilde g_s\ \ \ \ \ \ {\rm or }\
\ \ \ \ \  l_p^3=(\tilde l_s)^3 \tilde g_s .
\eqno(2.6)$$
The first relation in the above equation together with equation
(2.5) implies that ${R_{11}\over l_p^9}={1\over \tilde l_p^8}$. Equations
(2.5) and (2.6) are the known relations between the IIA theory and the so
called eleven dimensional M theory.  They encouraged the idea that eleven
dimensional M theory is the strong 
 coupling limit of the IIA string theory [10,11]. 
\medskip
{\bf {2.2 The correspondence between the eleven dimensional and  IIB
theories}}
\medskip
We now find the analogous relations between the fields, which 
 are associated with their Cartan sub-algebra, of the 
the eleven dimensional and  IIB
theories.  
Equating the eleven dimensional and IIB group elements of equation (1.6)
and equation (1.14) we find that 
$$g=\hat g\ \ \ \  {\rm or }\ \ \  \exp (\sum_{a=1}^{11} h_a{}^a K^a{}_a)=
\exp (\sum_{a=1}^{10} \hat h_a{}^a \hat K^a{}_a)\exp(\hat  A\hat R)
\eqno(2.7)$$
which using the identifications of equations (1.17) implies that 
$$h_a{}^a=\hat h_a{}^a +{1\over 3}\hat h_{10}{}^{10}, 
\ h_{10}{}^{10}=-{2\over 3}\hat h_{10}{}^{10}-{1\over 2}\hat A,\ 
h_{11}{}^{11}=-{2\over 3}\hat h_{10}{}^{10}+{1\over 2}\hat A
\eqno(2.8)$$
These relations hold without compactifications, but for a torus
compactification it is appropriate to adopt the  variables  
$$
\hat h_a{}^a=\ln {\hat  R_a\over \hat l_p},\ a=1,\dots , 10,\ \ 
\hat A=\ln \hat g_s
\eqno(2.9)$$
where $\hat l_p$ is the Planck length in the IIB theory and
$\hat g_s$ its string coupling. Introducing the IIB string scale by
$(\hat l_p)^8=\hat g_s^2 (\hat l_s)^8$ the relations given in equation
(2.8) become 
$${\hat l_s^4 \hat g_s\over l_p^3}=\hat R_{10},\ \ 
\ {R_{10}^6\over l_p^6}={\hat l_s^4\over\hat R_{10}^4 \hat g_s^2},\ \ 
{R_{11}^6\over l_p^6}={\hat l_s^4 \hat g_s^4\over \hat R_{10}^4 }
\eqno(2.10)$$
respectively. These are equivalent to the more familiar relations 
$$\hat g_s={R_{11}\over R_{10}},\ \ \hat l_s^2={l_p^3\over R_{11}},\ \ 
\hat R_{10}={l_p^3\over R_{10} R_{11}}
\eqno(2.11)$$
which relate the eleven dimensional theory reduced on rectangular torus
with radii $R_{10}$ and $R_{11}$ to the IIB theory reduced on a circle of
radius 
$\hat R_{10}$ [9,12,13]. 
\par
As explained in reference [7], there is a one to one map between all the
fields of the IIB and the eleven dimensional non-linearly realised
theories and not just those  associated with the Cartan
sub-algebra. We close this section by giving a
simple illustration of how this map works for a field outside the Cartan
sub-algebra. Equations (1.9) and (1.16) state that
$E_{10}=K^{10}{}_{11}=\hat R_2$ and as explained at the beginning of this
section this implies that the eleven dimensional field $h_{10}{}^{11}$
corresponds to the axion field $\hat \chi$ of the IIB theory. We now
enlarge the fields which are non-zero by including these fields in
addition to those associated with the Cartan sub-algebra. As a result,  
the eleven dimensional group element takes the form
$$g=\exp(\sum_{a=1}^{11} h_a{}^a K^a{}_a)\exp
(h_{10}{}^{11} K^{10}{}_{11}). 
\eqno(2.12)$$
Putting only the Cartan sub-algebra elements in the first exponential
will allow us to perform the computation more easily, but it  is not quite
the form given in the non-linear realisation of references [2,16] and 
used to find the eleven dimensional supergravity theory.  As a
result, we must use the form of the vierbein that follows from the $g$
of  equation (2.12); its non-vanishing components are    given by 
$$
e_\mu{}^a= \delta_\mu^a, \ a,\mu=1,\ldots ,9,\ 
e_\mu{}^a=\left( \matrix{e^{h_{10}{}^{10}}& e^{h_{10}{}^{10}}
h_{10}{}^{11}\cr 0& e^{h_{11}{}^{11}}\cr}\right)_\mu^{\ \ a} , 
a,\mu=10,11
\eqno(2.13)$$
On the other hand, the IIB group element can be written as 
$$
\hat g =\exp (\sum_{a=1}^{10} \hat h_a{}^a \hat K^a{}_a)
\exp(\hat  \chi \hat R_2) 
\exp(\hat  A\hat R)= \exp (\sum_{a=1}^{10} \hat h_a{}^a \hat K^a{}_a)
\exp(\hat  A\hat R) \exp(e^{\hat A}\hat \chi \hat R_2) 
\eqno(2.14)$$
The first  form of $\hat g$ is the one used to construct the non-linear
realisation of IIB supergravity in [6] while the second form is suitable
for our comparison with  eleven dimensional group element. To change
from one form to the other we used the relation $[\hat R,\hat
R_2]=-{1\over 2}[H_{10}, E_{10}]=-E_{10}=-\hat R_2$. 
\par
Setting $g=\hat g$ and using equations (1.17), (1.9) and (1.16) we find
the same relations of equation (2.8) as well as 
$$
e^{\hat A}\hat   \chi =h_{10}{}^{11}
\eqno(2.15)$$
\par
Let us now suppose that the ten and eleven directions of the eleven
dimensional theory are a torus with lengths $R_{10}$ and $R_{11}$. 
To discuss the properties of the torus it is simplest to make 
a rigid coordinate transformation from the coordinates $x^T=(x^{10},
x^{11})$ to the coordinates $y^T=(y^{10},y^{11})$ that diagonalises the
metric in these directions. In particular, we will 
diagonalise  the  veirbein  in the ten and eleven
directions.  We  denoted the latter by the matrix
$e$ which can be read off from the last relation in equation (2.13). The
transformation
$e\to \Lambda e$ given by 
$$\Lambda =\left( \matrix {1&m\cr
0&1}\right) ,
\eqno(2.16)$$
where $m=-{e_{10}{}^{11}\over e_{11}{}^{11}}$, has the desired result. The
new veirbein has the same diagonal components as the old one. Using
equation (2.13)  we find that
$m=-h_{10}{}^{11}{\exp( h_{10}{}^{10}}- h_{11}{}^{11})$.  In the diagonal
coordinates $y$ we take the cycles  of the torus to be given by 
$y^{11}=u,\  y^{10}=0;\ \  0\le u<1$ and
$y^{10}=v,\ y^{11}=0;\ \  0\le v<1$. The coordinate and parameterisation
invariant length of the first   cycle is 
$\int_0^1 e_{11}{}^{11} {d y^{11}\over du} du=e_{11}{}^{11}=R_{11}$  and
similarly with the invariant length of the second cycle is given by 
$e_{10}{}^{10}=R_{10}$. 
Hence,  we still have the relation ${R_{11}\over R_{10}}=\exp(\hat A)=\hat
g_s$ of equation (2.11). 
\par
In terms of the original $x$ coordinates  which are related by
$x=\Lambda^T y$ the cycles of the torus are 
$x^{10}=0,\ x^{11}=u;\ \  0\le u<1$ and 
 $x^{10}=v,\ x^{11}=mv;\ \ 0\le v<1$. If we define the complex coordinate 
$z=x^{11}-ix^{10}$ then the periods corresponding to the first and second
cycles are $z\to z=1$ and $z\to z+\tau$ respectively where
$\tau=\tau_1+i\tau_2$  with $\tau_1=1$ and $\tau_2=-m$.  Hence, the
modulus parameter of the torus  is given by 
$$
{\tau_1\over \tau_2}=m=\hat \chi
\eqno(2.17)$$
This agrees with the identification of references [12] and [13] after one
takes into account that one 
must make  the field redefinition  $\hat \chi \to
\exp(-\hat A)\hat \chi$ to find the $\hat \chi$ of [6] from that of [12]
in order to gain agreement between  the field equations of the
two references.  By a judicious choice of
coordinates we can, as in [12], arrange for 
$\tau_2$ to be $\exp(-\hat A)$, but the physically relevant quantity 
${\tau_1\over \tau_2}$ remains the same. 
\medskip
{\bf {3. Weyl transformations in the IIB and IIA theories }}
\medskip
The Weyl reflection $S_a$ corresponding to the simple root $\alpha_a$ on
any weight
$\beta$ is given by 
$S_a\beta =\beta-2{(\beta,\alpha_a)\over (\alpha_a,\alpha_a)}\alpha_a
$. For the simple roots this becomes 
$$
S_a\alpha_b =\alpha_b-2{(\alpha_b,\alpha_a)\over
(\alpha_a,\alpha_a)}\alpha_a=(s_a)_b{}^c\alpha_c
\eqno(3.1)$$
The action of the Weyl transformation $S_a$ on the Cartan sub-algebra of a
Kac-Moody algebra is given by 
$$ H^\prime_b=S_a H_b= (s_a)_b{}^c H_c
\eqno(3.2)$$
Since the Weyl group acts on Cartan
sub-algebra generators to give Cartan sub-algebra generators it makes
sense to consider their action on elements restricted to be of the form
of equation (1.6). Writing the group element
in  matrix form 
$g=\exp (q^T H)$, we conclude that the Weyl group acts on the fields $q$
as $q^{\prime T}=q^T s$, or $q^{\prime }=s^T q$ as $s^2=I$. Clearly,
these  transformations hold for the eleven dimensional
theory and the IIA and IIB theories. 
\par
To find the physical effects of the Weyl transformations we need to find
their action on  the physical variables $h_a{}^a$ and also 
the dilaton field, 
for the cases of the IIA and IIB theories,   However, 
the relationship between the Chevalley generators $H_a$ and the physical
generators  depends upon which theory we are considering and so the effect
of the Weyl transformations on the physical generators and fields is
different for each theory. Using matrix notation, in  the eleven
dimensional theory  we may write 
$H=\rho K$ and then the effect of the Weyl transformation is 
$S_a K=K^\prime =\rho^{-1} s_a\rho K= r_a K$ and so the physical fields
$h$ transform as
$h^\prime =r_a^T h$. However, for the IIB theory, $H=\hat \rho \hat K$ and
so  we have the equations 
$$ 
S_a \hat K=\hat K^\prime
=\hat \rho^{-1} s_a\hat \rho \hat K=\hat r_a \hat K,\ {\rm and } \ 
\hat h^\prime =\hat r_a^T\hat  h
\eqno(3.3)$$
The equation for IIA being found by replacing $\hat {}$ 's by $\tilde
{}$ 's. 
Using equations (2.3), (2.4) and (2.9) the effects  of the Weyl
transformations can then be readily deduced on the radii of any
compactified directions and the appropriate length scales and coupling
constants.   
\par
This calculation was carried out in reference [15] for the eleven
dimensional theory and we briefly summarize the result. The Weyl
transformations $S_a,\ a=1, \ldots ,10$ implied that 
$R_a\leftrightarrow R_{a+1}$, $l_p\to l_p$. However, $S_{11}$ induces
the transformations $h_a{}^{\prime a}=h_a{}^a+{1\over
3}(h_{9}{}^9+h_{10}{}^{10}+h_{11}{}^{11}),\ a=1,\ldots , 8$ and 
$h_a{}^{\prime a}=h_a{}^a
-{2\over 3}(h_{9}{}^9+h_{10}{}^{10}+h_{11}{}^{11}),\
a=9,10,11$ which in turn implies that 
$$
R_{9}^\prime ={l_p^3\over R_{10} R_{11}},\ 
R_{10}^\prime ={l_p^3\over R_{11} R_{9}},\ 
R_{11}^\prime ={l_p^3\over R_{9} R_{10}},\ 
\left(l_{p}^\prime\right)^3 ={l_p^6\over R_{9} R_{10} R_{11}}
\eqno(3.4)$$
\par
For the IIB theory, the Weyl transformations $S_a,\
a=1,\ldots ,8$ correspond $ \hat  K^a{}_{a}\leftrightarrow 
\hat  K^{a+1}{}_{a+1}$, for $a=1,\ldots ,8$ as well as $\hat R\to \hat R$.
The effect on the variables of equation (2.9) is $\hat R_a\leftrightarrow
\hat R_{a+1}$ for 
$a=1,\ldots ,7$ as well as $\hat g_s\to \hat g_s$. 
The Weyl transformation  $S_{11}$ leaves $\hat R$ and all the $\hat 
K^a{}_{a}$  inert except for $\hat  K^9{}_{9}\leftrightarrow \hat
K^{10}{}_{10}$. The effect is to take $\hat
R_9\leftrightarrow\hat  R_{10}$ with all other variables being inert. This
is consistent with the node labeled  eleven being the last on the
gravity line of the IIB theory and one finds that all the Weyl
transformations corresponding to all points on the gravity line just
exchanges the corresponding radii. 
\par
The Weyl transformation $S_{10}$ acts on the Cartan sub-algebra as 
$H^\prime _{10}=-H_{10}$,  $H^\prime _{9}=H_9+H_{10}$ all other elements
being inert. Using equation (1.15) we find that these transformations
imply that 
$$
\hat R^\prime=-\hat R,\ \ \  \hat  K^{\prime a}{}_a= \hat  K^a{}_a
\eqno(3.5)$$
Using equations (2.9) and (3.3), the effect on the physical variables is
given by 
$$
\hat A^\prime=-\hat A,\ \ \  h^{\prime}_{ a}{}^a= h_{ a}{}^a
\eqno(3.6)$$
Which in turn implies that 
$$
\hat g_s^\prime= \hat {1\over\hat  g_s},\ \ \
 \hat R_a^\prime=\hat R_a, \ a=1,\ldots ,10,\ \ \ \hat l_s^{\prime 2}=\hat
g_s
\hat l_s^2. 
\eqno(3.7)$$
This is just the non-perturbative S-duality transformations of the IIB
theory which holds if the theory is compactified or not. This is to be
expected as the node labeled ten just leads to an SL(2,{\bf R}) 
transformation of the supergravity theory. We note that in the eleven
dimensional theory,  node eleven is the last
node in the gravity line of this theory and the corresponding Weyl
transformation  swops the  eleventh and tenth coordinates. 
\par
Finally, we consider the Weyl transformation $S_9$ which induces the
transformations 
$H^\prime _{9}=-H_{9}$,  $H^\prime _{10}=H_9+H_{10}$, $H^\prime
_{8}=H_8+H_{9}$ with all other elements 
 of the Cartan
sub-algebra being inert. The transformation on the physical generators
is given by 
$$ \hat  K^{\prime a}{}_a= \hat  K^a{}_a, \ a=1,\ldots , 8,\ \ 
\hat  K^{\prime a}{}_a= \hat  K^a{}_a+{1\over 4}(\hat  K^{1}{}_{1}+\ldots
+\hat  K^{8}{}_{8})-{3\over 4}(\hat  K^{9}{}_{9}+\ldots 
+\hat  K^{10}{}_{10})-\hat R,\ 
a=9,10,  
$$
$$\hat R^\prime=\hat R+{1\over 8}(\hat  K^{1}{}_{1}+\ldots
+\hat  K^{8}{}_{8})-{3\over 8}(\hat  K^{9}{}_{9}+\ldots 
+\hat  K^{10}{}_{10})-{1\over 2}\hat R
\eqno(3.8)$$
\par
The corresponding effect  on the fields of the IIB theory is 
$$\hat h^\prime_a{}^a=\hat h_a{}^a+{1\over 4}(\hat
h_9{}^{9}+\hat h_{10}{}^{10}+{1\over 2}\hat A),\ a=1\ldots ,8
$$
$$
\hat h^\prime_a{}^a=\hat h_a{}^a-{3\over 4}(\hat
h_9{}^{9}+\hat h_{10}{}^{10}+{1\over 2}\hat A),\ a=9,10,\ \ 
\hat A^\prime=\hat A-(\hat
h_9{}^{9}+\hat h_{10}{}^{10}+{1\over 2}\hat A)
\eqno(3.9)$$
As a result the variables of equation (2.9) transform as 
$$
 \hat R_a^\prime=\hat R_a, \ a=1,\ldots ,8,\ 
{\hat R_a^\prime\over \hat R_a }={\hat l_s^2\over \hat R_9 \hat R_{10}},\ 
a=9,10,\ 
{\hat g_s^\prime\over \hat g_s}={\hat l_s^2\over \hat R_9 \hat R_{10}}
\eqno(3.10)$$
and $\hat l_s^\prime=\hat l_s$. 
We recognise this as a double T duality seen from the IIB
viewpoint. 
\par
We now briefly discuss the effect on the Weyl transformations of
$E_8^{+++}$ for the IIA theory.   The Weyl transformations $S_a,\
a=1,\ldots, 9$ takes $K^a{}_a\leftrightarrow K^{a+1}{}_{a+1}$ and so 
$R_a \leftrightarrow R_{a+1}$. The Weyl transformation $S_{11}$  
leads to the double T duality 
$$R_{a}^\prime=R_a,\  a=1,\ldots, 8;\ \ 
\tilde R_{9}^\prime ={\tilde l_s^2\over \tilde R_{10}};\ 
\tilde R_{10}^\prime ={\tilde l_s^2\over\tilde  R_{9}};\ 
\tilde g_s^\prime={\tilde g_s \tilde l_s^2\over \tilde R_{9} \tilde
R_{10}} ;\ \tilde l^\prime_s=\tilde l_s
\eqno(3.11)$$
Finally, the Weyl transformation $S_{10}$ induces the changes 
$$
\tilde h^\prime_a{}^a=\tilde h_a{}^a+{1\over 8}\tilde
h_{10}{}^{10}+{3\over 32}\tilde A,\ a=1\ldots ,9
$$
$$
\tilde h^\prime_{10}{}^{10}=\tilde h_{10}{}^{10}
-{7\over 8}\tilde h_{10}{}^{10}+{21\over 32}\tilde A,\
a=9,10,\ \ 
\tilde A^\prime=\tilde A+{3\over 2}\tilde h_{10}{}^{10}
-{9\over 8}\tilde A
\eqno(3.12)$$
which leads to 
$$
R_{a}^\prime=R_a,\  a=1,\ldots, 9;\ \ 
\tilde R_{10}^\prime =\tilde l_s\tilde g_s ;\ 
\tilde g_s^{\prime 2}=
\left({\tilde R_{10}\over  \tilde l_s}\right)^3{1\over
\tilde g_s};\ 
\left({\tilde l_s^\prime\over \tilde l_s}\right)^2=
{\tilde l_s\tilde g_s\over \tilde R_{10}}
 \eqno(3.13)$$
Clearly, this is a non-perturbative relation which is in some sense the
IIA analogue of the SL(2,{\bf Z}) symmetry of the IIB theory. 
\medskip  
{\bf {3. Discussion }}
\medskip
One could use the same techniques as used in this paper to identify
the relations between other $G^{+++}$ non-linearly realised theories where
there  is a choice of gravity sub-algebra. 
\par
The eleven dimensional, IIA and
IIB theories are all expected to possess a non-linearly realised
$E_8^{+++}$ symmetry [2,6].  Although, their differences arise from the
way their 
 gravity sub-algebras are embedded,  their common symmetry allows one
to establish a one to one correspondence between the fields of these
theories [7]. In this paper,  we have found the simplest consequences
of this correspondence which are those for the fields associated with the
Cartan sub-algebra of $E_8^{+++}$. We have recover the known relations
[8,9,10,11] between the three theories. We also gave one example of the
correspondence for a field outside the Cartan sub-algebra and recovered
the fact [12,13] that the axion field of the IIB theory dimensionally
reduced on a circle can be identified with the modulus of the 
 two dimensional torus used to dimensionally reduce the eleven
dimensional theory. 
\par
The correspondence between the three theories resulting from   their
common
$E_8^{+++}$ symmetry  implies many more results, such as the eleven
dimensional origin of the massive IIA theory and the IIB space-filling
brane [7]. However, the purpose of this paper is to demonstrate that the
underlying $E_8^{+++}$ symmetry  can be used to  find results central to
string theory in a very simple way. 
\par
As we noted above, the identifications of the fields of the three theories
should hold even if  one does not perform  a dimensional reduction.
In this case one is the fields which depend on the
generalised coordinates [17] of the theory and, as explained in reference
[7],  one must  then also  swop  the generalised
coordinates of the theory. However, these  includes central charge
coordinates as well as the usual coordinates of space-time and their 
interchange will have far reaching  effects on the theory. 
\par
We also computed the effect of the Weyl transformations of the IIA and
IIB $E_8^{+++}$ theories on the diagonal components of the metric and
dilaton to recover the expected U-duality symmetries of these theories. 
It would be interesting to compare these results with the different
perturbative  sub-algebras  of the $E_8^{+++}$ algebra for the IIA and
IIB theories found  in [20]. 
\medskip 
{\bf Acknowledgments}
I wish to thank Matthias Gaberdiel  and Dominic Clancy for useful
discussions.  I wish also to thank the Erwin Schršdinger International
Institute for Mathematical Physics at Wien and the Department of Physics
at Heraklion for their hospitality.   This research was supported by a
PPARC senior fellowship PPA/Y/S/2002/001/44 and  in part by the PPARC
grants  PPA/G/O/2000/00451,  PPA/G/S4/1998/00613 and the EU Marie Curie,
research training network grant HPRN-CT-2000-00122. 
\medskip
{\bf References}
\medskip
\item{[1]} E. Cremmer, B. Julia and J. Scherk, 
{\it  Supergravity theory in eleven dimensions},  Phys. Lett. {\bf 76B}
(1978) 409.
 \item{[2]} P. West, {\sl $E_{11}$ and M Theory}, Class. Quant.
Grav. {\bf 18 } (2001) 4443, {\tt hep-th/0104081}
\item{[3]} I.C.G. Campbell and  P. West, {\it N=2 d=10 nonchiral
supergravity
     and its spontaneous compactifications}, Nucl. Phys. {\bf B243} (1984),
     112; M. Huq, M. Namanzie, {\it Kaluza-Klein supergravity in ten
     dimensions}, Class. Quant. Grav. {\bf 2} (1985); F. Giani, M. Pernici,
     {\it N=2 supergravity in ten dimensions}, Phys. Rev. {\bf D30} (1984),
     325
\item{[4]} J. Schwarz and  P. West {\it Symmetries and Transformations of
     chiral N=2, D=10 supergravity}, Phys. Lett. {\bf B126} (1983), 301.
\item{[5]}
     J. Schwarz, {\it Covariant field equations of chiral N=2 D=10
     supergravity}, Nucl. Phys. {\bf B226} (1983), 269; P. Howe and  P.
West,
     {\it The complete N=2, d=10 supergravity}, Nucl. Phys. {\bf B238}
     (1984), 181
\item{[6]} I. Schnakenburg and P. West, {\sl Kac-Moody Symmetries of
IIB supergravity}, Phys. Lett. {\bf B 517} (2001) 137-145, {\tt
hep-th/0107181} 
\item{[7]} P. West, {\it  The  IIA, IIB and eleven dimensional 
theories and their common $E_{11}$ origin }, hep-th/0402140. 
\item{[8]} J. Dai, R. Leigh and J. Polchinski, {\it New connections
between string theories},  Mod. Phys. Lett. {\bf A4}  (1989) 2073. 
\item{[9]} M. Dine, P. Huet and N. Seiberg, 
{\it Large and small radius in string theory}, 
 Nucl. Phys. {\bf B322} (1989), 301. 
\item{[10]} P. Townsend, {\it The eleven-dimensional supermembrane
revisited}, Phys. Lett. {\bf B 350} (1995) 184, hep-th/9501068. 
\item{[11]} E. Witten, {\it String theory dynamics in various
dimensions},  Nucl. Phys. {\bf B443} (19995) 85, hep-th/9503124. 
\item{[12]} J. Schwarz, {\it The Power of M Theory}, Phys.Lett. B367
(1996) 97-103, hep-th/9510086; {\it An SL(2,Z) Multiplet of Type IIB
Superstrings},  Phys.Lett. {\bf B360} (1995) 13-18; Erratum, B364 (1995)
252,  hep-th/9508143. 
\item{[13]} P. Aspinwall, {\it Some relations between dualities in
string theories}, Nucl. Phys. Proc Suppl. {\bf 46} (1996) 30,
hep-th/9508154. 
\item{[14]}  M. R. Gaberdiel, D. I. Olive and P. West, {\sl A
class of Lorentzian Kac-Moody algebras}, Nucl. Phys. {\bf B 645}
(2002) 403-437,  hep-th/0205068. 
\item{[15]}  F. Englert, L. Houart, A. Taormina and P. West,
{\sl The Symmetry of M-theories}, JHEP {\bf 0309} (2003) 020, 
 hep-th/0304206. 
\item{[16]} P.~C. West, {\sl Hidden superconformal symmetry in {M}
    theory },  JHEP {\bf 08} (2000) 007, {\tt hep-th/0005270}
\item{[17]} P. West, {\sl $E_{11}$, SL(32) and Central Charges},
Phys. Lett. {\bf B 575} (2003) 333-342, {\tt hep-th/0307098}
\item{[18]} C.M. Hull and P.K. Townsend, {\it Unity of  
superstring  dualities}, Nucl. Phys. {\bf B 438}
(1995) 109. hep-th/9410167. 
\item{[19]} A. Kleinschmidt and  P. West, {\sl Representations of ${\cal
G}^{+++}$ and the role of space-time}, JHEP {\bf 0402} (2004) 033,
hep-th/0312247. 
\item{[20]} M. Gaberdiel and  P. West, {\it  Kac-Moody algebras in
perturbative string theory},  JHEP {\bf 0208} (2002) 049, hep-th/0207032 Ê

\end